\documentclass[12pt]{iopart}
\usepackage{iopams} 
\usepackage{cite}    
\usepackage{graphicx}

\begin{document} 
 
\title[Plasmonic light scattering in graphene-coated subwavelength wires]
{Tunable plasmonic enhancement of light scattering and absorption in graphene-coated subwavelength wires}
\author{M\'aximo Riso$^{1}$, Mauro Cuevas$^{2}$ and Ricardo A. Depine$^{1}$}
\address{$^{1}$ Grupo de Electromagnetismo Aplicado, Departamento de F\'{\i}sica, FCEN, Universidad de Buenos Aires and IFIBA, Consejo Nacional de Investigaciones Cient\'{\i}ficas y T\'{e}cnicas (CONICET), Ciudad Universitaria, Pabell\'{o}n I, C1428EHA, Buenos Aires, Argentina}

\address{$^{2}$ Facultad de Ingenier\'ia y Tecnolog\'ia Inform\'atica, Universidad de Belgrano, Villanueva 1324, C1426BMJ, Buenos Aires, Argentina and Consejo Nacional de Investigaciones Cient\'{\i}ficas y T\'{e}cnicas (CONICET)}

\ead{rdep@df.uba.ar}

\begin{abstract} 
The electromagnetic response of subwavelength wires coated with a graphene monolayer illuminated by a linearly polarized plane waves is investigated. The results show that the scattering and extintion cross-sections of the coated wire can be dramatically enhanced when the incident radiation resonantly excites localized surface plasmons. The enhancements
occur for p--polarized incident waves and for excitation frequencies that correspond to complex poles in the coefficients of the multipole expansion for the scattered field. By dynamically tuning the chemical potential of graphene, the spectral position of the enhancements can be chosen over a wide range. 
\end{abstract} 

\section{Introduction} \label{intro}

Advanced light manipulation relies heavily upon controlling the fundamental processes of light-matter interaction. 
This paper is concerned with light scattering, a process which plays a key role in current light control applications like 
light trapping \cite{trapping1}, Anderson localization \cite{anderson1,anderson2} or photonic band gaps \cite{pbg1}.  
%
%
Particularly, light scattering by subwavelength particles has received considerable attention in relation to 
\emph{plasmonic} systems, where some exciting experimental and theoretical developments
can be produced when incident radiation is coupled to the collective electron charge oscillations known as localized surface plasmons \cite{review1,SPP1,localized1,fluorescence1,raman1}. 
%
Excitation of localized surface plasmons 
by an incident electromagnetic field at the frequency where resonance occurs results in an enhancement of the local electromagnetic field as well as in the appearance of intense absorption bands. Moreover, the efficiency of 
light scattering, determined by the strength of the electric field in the material, is also enhanced under resonant excitation of surface plasmons. 
The fact that the details of the resonance are highly sensitive to particle characteristics such as material,  
size, shape and enviroment constitutes the precise property which has prompted an exponential growth in the exploitation of localized surface plasmons in nanoscale optics and photonics \cite{localized1}, giving applications such as molecular-recognition elements \cite{fluorescence1}, amplifiers \cite{raman1} and biosensors.  
Applications and new developments in the field of light scattering by plasmonic particles have been recently 
reviewed in \cite{review1}. 
%

Natural plasmonic materials such as metals or metal-like materials are nonmagnetic (magnetic permeability of free--space) and have electric permittivity with a negative real part. 
Currently available plasmonic materials have a number of shortcomings, most notably power losses 
and bad tunability, which limit the further development of plasmonics. The recent advent of graphene --a monolayer of carbon atoms arranged in a hexagonal lattice-- have just meet the need of surface plasmons since it offers relatively low loss, high confinement and good tunability, three characteristics that make graphene a promising plasmonic  material alternative to 
noble metals \cite{SPmaterials1,SPmaterials2,SPmaterials3,SPmaterials4}. 
The waveguiding characteristics of graphene plasmons have been investigated in different planar structures, 
such as infinite sheets \cite{sheet}, ribbons \cite{rib1,rib2} or grooves \cite{groove}. 
Since the wave vector needed to resonantly excite these plasmons is always greater than that available to incident 
electromagnetic radiation, the electromagnetic response of graphene surface plasmons has been investigated for 
phase matching configurations such as the prism \cite{ATR1} and grating \cite{G1,G2,G3} couplers 
as well as for layered media \cite{lay1,lay2}. 

Quite recently non-planar configurations \cite{cil1,cil2,cil3,cil4,scatsphere} have also begun to attract particular 
attention since cylindrical and spherical structures have been experimentally constructed for a number of 
applications \cite{exp1,exp2,exp3,exp4,expesf1,expesf2}. 
The electromagnetic response and the plasmonic properties of graphene--coated nanospheres have been recently examined in \cite{scatsphere} and 
the modal characteristics of surface plasmons propagating along the axis of graphene--coated wires have been studied in 
\cite{cil1,cil2,cil3,cil4}. However, the resonant excitation of {\em localized} plasmon modes in graphene-coated wires
has not been investigated yet.
The primary motivation of the work described here is to study the plasmonic response of graphene--coated 2D particles (wires). 
For metallic or metallic-like substrates, the 
coating is expected to modify the localized surface plasmons already existing in the bare particle. On the other hand, for substrates made of nonplasmonic materials the 
coating is expected to introduce localized surface plasmons mechanisms which were absent in the bare particle. In both cases, the good tunability of the graphene coating is expected to lead to unprecedented control over the location and magnitude of the particle plasmonic resonances. In this paper we study the scattering of electromagnetic waves normally incident on graphene--coated cylinders with circular cross--section, and present results for both dielectric and metallic substrates. 

The plan of the paper is as follows. In Section \ref{teoria} we sketch an analytical method of scattering based on the separation of variables approach and obtain a solution for the scattered fields in the form of an infinite series of cylindrical multipole partial waves. 
In Section \ref{resultados} we give examples of scattering and extinction cross--section corresponding to subwavelength wires tightly coated with a graphene layer. Both transparent and metallic susbstrates are considered. In both cases, great enhancements are observed for p--polarized incident radiation and for excitation frequencies 
corresponding to localized plasmons of the graphene-coated cylinder. 
%
Finally, in Section \ref{conclusiones} we summarize and discuss the results obtained. The Gaussian system of units is used and an $\exp(-i\omega t)$ time--dependence is implicit throughout the paper, with $\omega$ the angular frequency, $t$ the time, and $i=\sqrt{-1}$. 	

\section{Theory} \label{teoria} 

\begin{figure}[htbp!]
\centering
\resizebox{0.60\textwidth}{!}
{\includegraphics{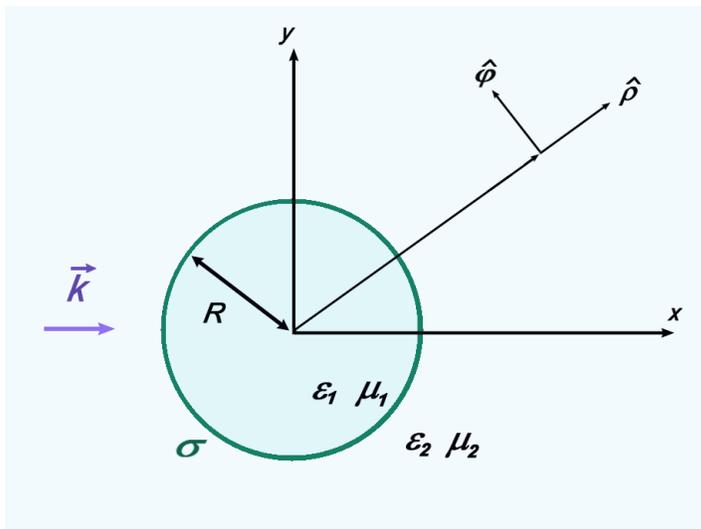}} 
\caption{Schematic of the scattering problem}
\label{dibujo}
\end{figure}

We consider a graphene--coated cylinder with circular cross--section (radius $R$) centered at $x$=0, $y$=0 (see Figure \ref{dibujo}). The wire substrate may be dielectric or conducting (electric permittivity $\varepsilon_{1}$ and magnetic permeability $\mu_{1}$) and the coated wire is embedded in a transparent medium (electric permittivity $\varepsilon_{2}$ and magnetic permeability $\mu_{2}$). The graphene layer is treated as an infinitesimaly thin, local and isotropic two-sided layer with surface conductivity $\sigma(\omega)$ given by the Kubo formula \cite{kubo1,kubo2}. 
The wave vector of the incident radiation is directed along $\hat x$. In this case, the scattering problem can be handled in a scalar way since the solution to any incident polarization can be described as a linear combination of the solutions obtained in two fundamental scalar problems: electric field parallel to the main section of the cylindrical surface (p polarization or $E_z$=0 modes) and magnetic field parallel to the main section of the cylindrical surface (s polarization or $H_z$=0 modes). 

To obtain analytical solutions to the scattering problem we closely follow the usual separation of variables approach \cite{libro1,libro2}. In a first step, the non-zero components of the total electromagnetic field along the axis of the cylinder for each polarization case, denoted by $F (\rho,\varphi)$, are expanded as series of cylindrical harmonics, one for the internal region ($\rho < R$, subscript 1) and another one for the external region ($\rho > R$, subscript 2). 
When the incident electric field is contained in the $x-y$ plane (p polarization), the expansions for the total 
magnetic field along the axis of the cylinder are 
\begin{eqnarray}
&& F_{1}(\rho,\varphi)= \sum_{n = -\infty}^{\infty} c_{n} J_n(k_1\rho)\,\exp{i n\varphi}  \,, \label{Fz1p} \\ 
&& F_{2}(\rho,\varphi)= \sum_{n = -\infty}^{\infty} \left [ A_{0} \,i^{n} J_n(k_2\rho) +  a_{n}\;H_n^{(1)}(k_2\rho)
\right ]\,\exp{in\varphi}\,, \label{Fz2p}
\end{eqnarray}
where $a_{n}$ and $c_{n}$ are unknown complex coefficients, 
$k_{j}=\frac{\omega}{c}\sqrt{\varepsilon_{j}\mu_{j}}$ ($j=1, 2$), $c$ is the speed of light in vacuum, $A_{0}$ is the amplitude of the incident magnetic field (parallel to $\hat z$) and $J_n$ and $H_n^{(1)}$ are the $n$-th Bessel and Hankel functions of the first kind respectively.  
In the same manner, when the incident magnetic field is contained in the $x-y$ plane (s polarization), the expansions for the total electric field along the axis of the cylinder are 
\begin{eqnarray}
&& F_{1}(\rho,\varphi)= \sum_{n = -\infty}^{\infty} d_{n} J_n(k_1\rho)\,\exp{i n\varphi}  \,, \label{Fz1s} \\ 
&& F_{2}(\rho,\varphi)= \sum_{n = -\infty}^{\infty} \left [ B_{0} \,i^{n} J_n(k_2\rho) +  b_{n}\;H_n^{(1)}(k_2\rho)
\right ]\,\exp{in\varphi}\,, \label{Fz2s}
\end{eqnarray}
where $B_{0}$ is the amplitude of the incident electric field (parallel to $\hat z$) and $b_{n}$ and $d_{n}$ unknown complex coefficients. 

In a second step, the boundary conditions at $\rho = R$ are invoked to obtain the unknown coefficients $a_{n}$, $b_{n}$, $c_{n}$ and $d_{n}$ in terms of the incident amplitudes $A_{0}$ and $B_{0}$.  
The boundary conditions for the graphene--coated cylinder are: i) the tangential component of the total electric field $\vec E$ is continuous and ii) the discontinuity of the tangential component of the total magnetic field $\vec H$ is proportional in magnitude to the surface current density (that is, to the graphene surface conductivity).  
%
%
In terms of $F$, the boundary conditions at $\rho = R$ can be expressed as 
\begin{eqnarray}
\frac{1}{\varepsilon_{2}} \frac{\partial F_{2}}{\partial \rho} = 
\frac{1}{\varepsilon_{1}} \frac{\partial F_{1}}{\partial \rho} \,\,\,\,\,\,\,\,\,\,\,\,  \mbox{and} \,\,\,\,\,\, \,\,\,\,\,\, 
F_{2} - F_{1}= \frac{4i\pi}{\omega \varepsilon_{1}}\sigma \frac{\partial F_{1}}{\partial \rho}\,, \label{contornop}
\end{eqnarray}
for p polarization, and 
\begin{eqnarray}
F_{1}=F_{2}\,\,\,\,\,\, \,\,\,\,\,\, \mbox{and} \,\,\,\,\,\, \,\,\,\,\,\, 
\frac{1}{\mu_{2}} \frac{\partial F_{2}}{\partial \rho} - 
\frac{1}{\mu_{1}} \frac{\partial F_{1}}{\partial \rho} = -\frac{4i\pi \omega}{c^2}\sigma F_{1} \,, \label{contornos}
\end{eqnarray}
for s polarization. 
%
%
%
%
Finally, the amplitudes of the scattered field can be written as
%
\begin{equation}
\hspace*{-2.2cm} a_n=\frac{ -i^{n}\Bigl[\varepsilon_1k_2J_n(x_1)J_n'(x_2)-\varepsilon_2k_1J_n'(x_1)J_n(x_2)+ \frac{4\pi}{c}\sigma  \frac{c}{\omega} i  k_1k_2J_n'(x_1)J_n'(x_2)\Bigr]}
{\varepsilon_1k_2J_n(x_1)H_n'^{(1)}(x_2)-\varepsilon_2k_1J_n'(x_1)H_n^{(1)}(x_2) + \frac{4\pi}{c}\sigma  \frac{c}{\omega} i  k_1k_2J_n'(x_1)H_n'^{(1)}(x_2)}A_{0}, \label{an}
\end{equation}
\begin{equation}
\hspace*{-2.2cm} b_n=\frac{ -i^{n}\Bigl[\varepsilon_2k_1J_n(x_1)J_n'(x_2)-\varepsilon_1k_2J_n'(x_1)J_n(x_2)+ \frac{4\pi}{c}\sigma  \frac{c}{\omega} i  k_1k_2J_n(x_1)J_n(x_2)\Bigr]}
{\varepsilon_2k_1J_n(x_1)H_n'^{(1)}(x_2)-\varepsilon_1k_2J_n'(x_1)H_n^{(1)}(x_2) + \frac{4\pi}{c}\sigma  \frac{c}{\omega} i  k_1k_2J_n(x_1)H_n^{(1)}(x_2)}B_{0}\,, \label{bn}
\end{equation}
while the amplitudes of the fields inside the wire are given by 
\begin{equation}
\hspace*{-2.cm} c_n=\frac{\varepsilon_1k_2 i^{n}\Bigl[J_n(x_2)H_n'^{(1)}(x_2)-J_n'(x_2)H_n^{(1)}(x_2)\Bigr]}
{\varepsilon_1k_2J_n(x_1)H_n'^{(1)}(x_2)-\varepsilon_2k_1 J_n'(x_1)H_n^{(1)}(x_2) + \frac{4\pi}{c}\sigma  \frac{c}{\omega} i  k_1k_2J_n'(x_1)H_n'^{(1)}(x_2)}A_{0}\,,\label{cn}
\end{equation}
\begin{equation}
\hspace*{-2.cm} d_n=\frac{\varepsilon_2k_1 i^{n} \Bigl[J_n(x_2)H_n'^{(1)}(x_2)-J_n'(x_2)H_n^{(1)}(x_2)\Bigr]}
{\varepsilon_2k_1J_n(x_1)H_n'^{(1)}(x_2)-\varepsilon_1k_2J_n'(x_1)H_n^{(1)}(x_2) + \frac{4\pi}{c}\sigma  \frac{c}{\omega} i  k_1k_2J_n(x_1)H_n^{(1)}(x_2)}B_{0}\,, \label{dn}
\end{equation}
where $x_1=k_1R$, $x_2=k_2R$. 

As in the problem of scattering by a wire without a graphene coating, the amplitudes given by equations (\ref{an})-(\ref{dn}) allow us to obtain the electromagnetic field everywhere in space as well as other quantities of interest such as differential, absorption and extinction cross--sections (see \cite{libro1,libro2} for details). 
The multipole coefficients $a_n$ and $b_n$ for the scattered field have essentially the same form as those corresponding to a bare wire \cite{libro1,libro2}, except for additive corrections proportional to $\sigma$ in numerator and denominator. Similar additive corrections have been recently reported for graphene-coated spheres \cite{scatsphere}. 
%
%

\section{Resonant excitation of localized surface plasmons} \label{resultados} 
%

To simulate the electromagnetic response of graphene-coated cylinders we assume that $\sigma(\omega)$, the complex surface conductivity of graphene, is given by the high frequency expression derived from Kubo formula (equation (1), Ref. \cite{kubo1}),  
including interband and intraband transition contributions. 
Apart from the angular frequency $\omega$, the value of $\sigma$ depends on the chemical potential $\mu_c$ (controlled with the help of a gate voltage), the ambient temperature $T$ and the carriers scattering rate $\gamma_c$. 
For the intraband ($\sigma ^{intra}$) and interband ($\sigma ^{inter}$) contributions to $\sigma(\omega)$ we have used the following expressions \cite{kubo1,kubo2}
\begin{equation}
\hspace*{-1.4cm}      \sigma ^{intra}(\omega) =\frac{2ie^2T}
     {\pi\hbar(\omega+i\gamma_c)}
\ln{[2\cosh(\mu_c /2T)]}
 \label{sigm}    \, , 
 \end{equation}
\begin{equation}
\hspace*{-1.4cm}     \sigma^{inter}(\omega) =
    \frac{e^2}{4\hbar}\left[
\frac{1}{2}+\frac{1}{\pi}\arctan[(\omega-2\mu_c)/2T] 
-\frac{i}{2\pi}\ln
    \frac{(\omega+2\mu_c)^2}
{(\omega-2\mu_c)^2+(2T)^2}
     \right]\; , 
 \label{ibd} \end{equation}
where $\hbar$ is the reduced Planck constant and $e$ the elementary charge. 
In all the calculations we have used $T=300^\circ$ K and $\gamma_c=0.1$ meV. 

\begin{figure}[htbp!]
\centering
\resizebox{0.80\textwidth}{!}
{\includegraphics{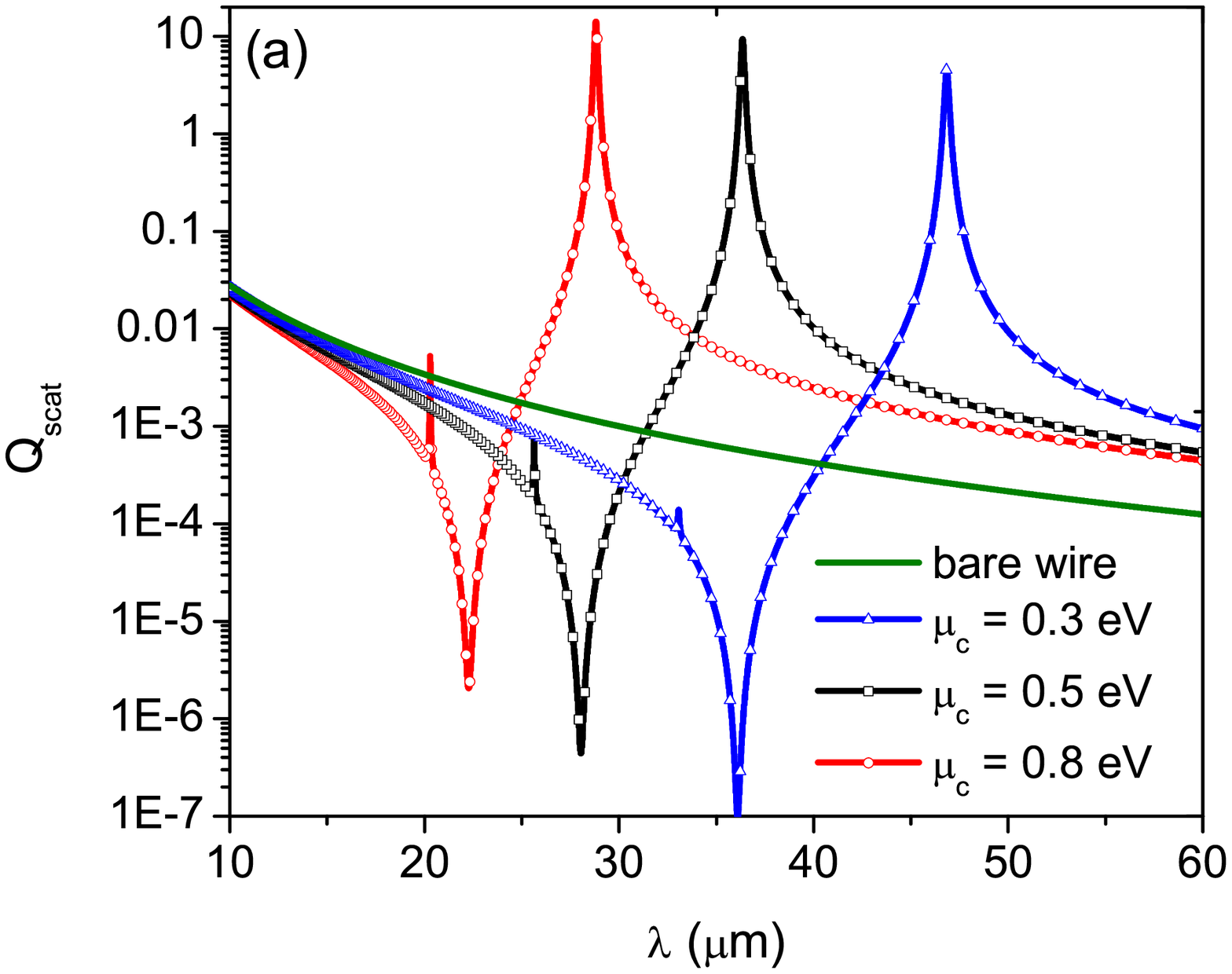}}
\resizebox{0.80\textwidth}{!}
{\includegraphics{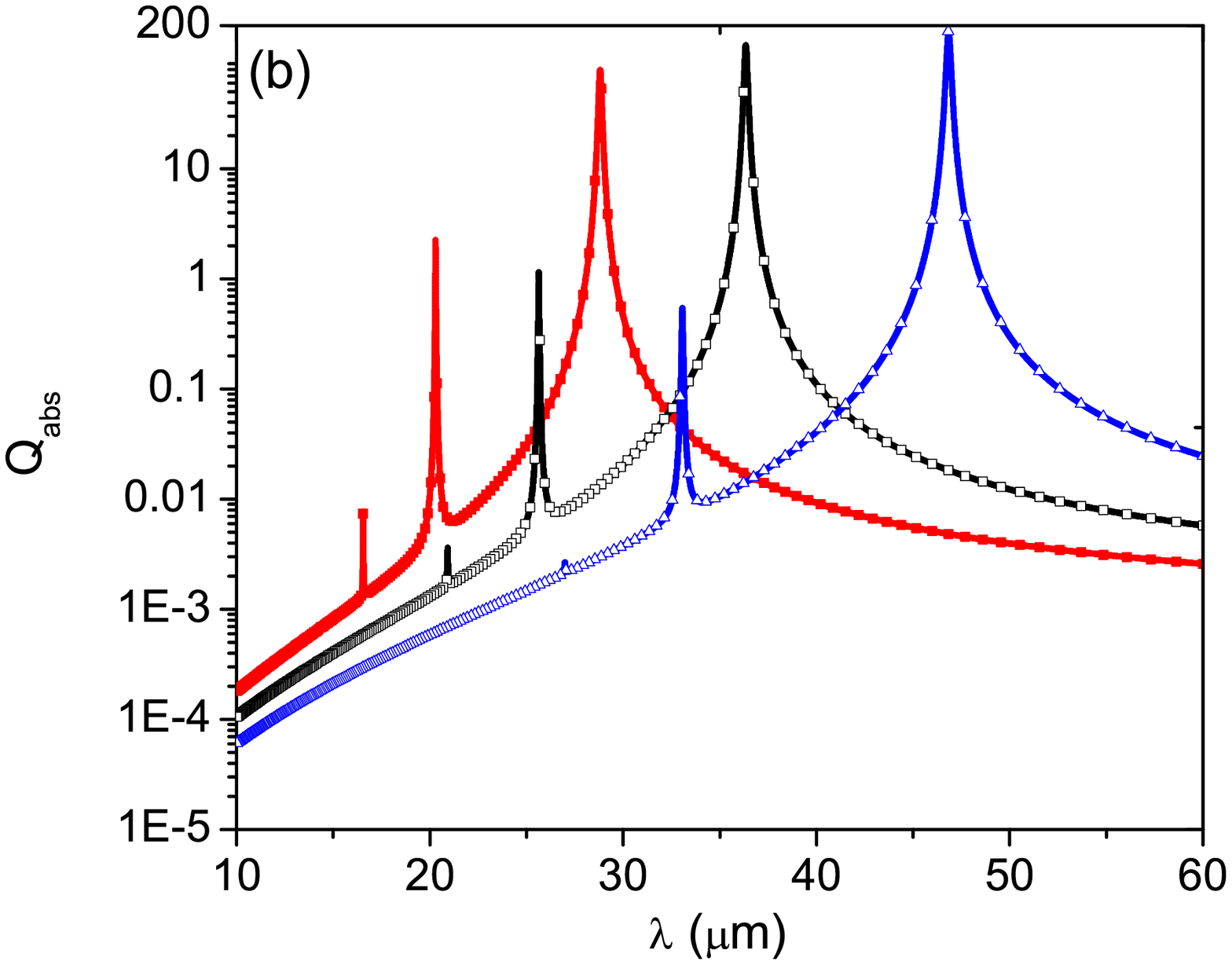}}
\caption{Scattering (a) and absorption (b) cross-section spectra for a graphene-coated cylinder illuminated by a p-polarized plane wave and for different values of the chemical potential $\mu_c$. The scattering cross-section curve corresponding to the uncoated cylinder is given as a reference. The incident wave is $p$-polarized, $R=0.5\;\mu$m, $\varepsilon_{1}=3.9$ and $\mu_{1}=\mu_{2}=\varepsilon_{2}=1$.}
\label{scat-absor-diel}
\end{figure}

In Figure \ref{scat-absor-diel} we plot the scattering and absorption cross-sections (per unit length) for a wire with a radius $R=0.5 \;\mu$m, made with a nonplasmonic, transparent material 
($\varepsilon_{1}=3.9$, $\mu_{1}=1$) in a vaccum ($\mu_{2}=\varepsilon_{2}=1$). 
The excitation frequencies are in the range between $5$ THz (incident wavelength $60 \;\mu m$) and $30$ THz (incident wavelength $10 \;\mu$m), the incident wave is p-polarized and different values of $\mu_c$ have been considered. 
In Figure \ref{scat-absor-diel}a the scattering cross-section spectrum of the bare wire (continuous line) is also given as a  reference. The absorption cross-section of the bare wire vanishes identically and is not shown in Figure \ref{scat-absor-diel}b. 
We see that while the scattering cross-section spectrum of the bare wire does not show any plasmonic feature in this spectral range, great enhancements in both the scattering and the absorption cross-section spectra occur at wavelengths near 
$46.84 \;\mu$m ($\mu_c=0.3$ eV), $36.35 \;\mu$m ($\mu_c=0.5$ eV) and $28.81 \;\mu$m ($\mu_c=0.8$ eV). 
At these spectral positions we obtain enhancement factors in the scattering cross-section of around four orders of magnitude. 
Other local maxima, particularly noticeable in the absorption cross-section spectra, occur at wavelengths near 
$46.84 \;\mu$m, $33.06 \;\mu$m (for $\mu_c=0.3$ eV), $36.35 \;\mu$m, $25.64 \;\mu$m, $20.92 \;\mu$m (for $\mu_c=0.5$ eV) and 
$28.81 \;\mu$m, $20.29 \;\mu$m, $16.55 \;\mu$m (for $\mu_c=0.8$ eV). 
The local maxima observed in the scattering and absorption cross-sections displayed in Figure \ref{scat-absor-diel} 
correspond to plasmonic resonances of the graphene-coated wire and can be obtained from the complex poles of the 
coefficients $a_n$ and $c_n$ of the multipole expansions 
for the electromagnetic field. The correspondence can be clearly seen by noting that the spectral positions of the peaks are in very good agreement with the real part of the complex wavelengths $\Lambda_n$ given in Table 1 ($n=1, \ldots 4$), with $\Lambda_n$ representing the complex root of the common denominator $g_n(\Lambda)$ in equations (\ref{an}) and (\ref{cn}), and $\Lambda=2\pi c/\omega$. For the numerical calculation of $\Lambda_n$ we have used a Newton-Raphson method and the same parameters used for curves in Figure~\ref{scat-absor-diel}. 
We observe that the scattering cross-sections in Figure \ref{scat-absor-diel}a display minima 
at wavelengths near $36.09 \;\mu$m ($\mu_c=0.3$ eV), $28.04 \;\mu$m ($\mu_c=0.5$ eV) and $22.27 \;\mu$m ($\mu_c=0.8$ eV). 
We found that the spectral position of these minima are in excellent agreement with the real part of the complex wavelengths where the numerator of the coefficient $a_1$ becomes zero. 
Taking into account that a minimization of the scattering cross-section is a necessary condition to obtain invisibility or transparency of an object, and that the scattering cross-section of the bare wire does not exhibit a minimum in this spectral range, the observation of these minima is relevant in the context of invisibility cloacks. 
These results, together with those shown in Figure \ref{scat-absor-diel}, illustrate a great advantage of the graphene coating, which provides unprecedented control over the spectral location of the resonances of the wire via changes in the chemical potential $\mu_c$, ultimately controlled through electrostatic gating. 

The correspondence between plasmonic resonances of the graphene coating and the enhancement of scattering and absorption in the coated wire is further evidenced when the cross-sections of the same systems considered in Figure \ref{scat-absor-diel} are calculated for s-polarized --instead of p-polarized-- illumination.  In this case (not shown), no enhancements in the 
cross-sections of the graphene-coated wire are observed, which is consistent with the fact that localized surface plasmons are not supported in this polarization mode, since the electric field in the graphene coating can only induce electric currents directed along the wire axis, and not along the azimuthal direction $\hat \varphi$, a necessary condition for the existence of localized surface plasmons in the graphene circular cylinder. 
%
%
\begin{table}
\begin{center}
\begin{tabular}{|c|c|c|c|c|}
\hline
$n$ &   $\mu_c=0.3$ eV  & $\mu_c=0.5$ eV &   $\mu_c=0.8$ eV    \\
\hline
$1$ &  $46.8378+i0.1216$ & $36.3464+i0.0961$   &  $28.8092+i0.0875$   \\
$2$ &  $33.0651+i0.0439$ & $25.6361+i0.0264$   &  $20.2956+i0.0167$    \\
$3$ &  $26.9865+i0.0292$ & $20.9175+i0.0175$   &  $16.5531+i0.0109$   \\
$4$ &  $23.3660+i0.0219$ & $18.1086+i0.0131$   &  $14.3272+i0.0082$    \\
\hline
\end{tabular}

\vspace{0.4cm}

Table 1: Complex roots $\Lambda_n$ of $g_n(\Lambda)=0$ in , where $g_n$ is the common denominator of the coefficients $a_n$ 
and $c_n$ and $\Lambda=2\pi c/\omega$. The parameters correspond to those used to calculate the scattering and absorption cross-sections shown in Figure~\ref{scat-absor-diel}. 

\end{center}
%
%
\end{table}

\begin{figure}[htbp!]
\centering
\resizebox{0.32\textheight}{!}
{\includegraphics{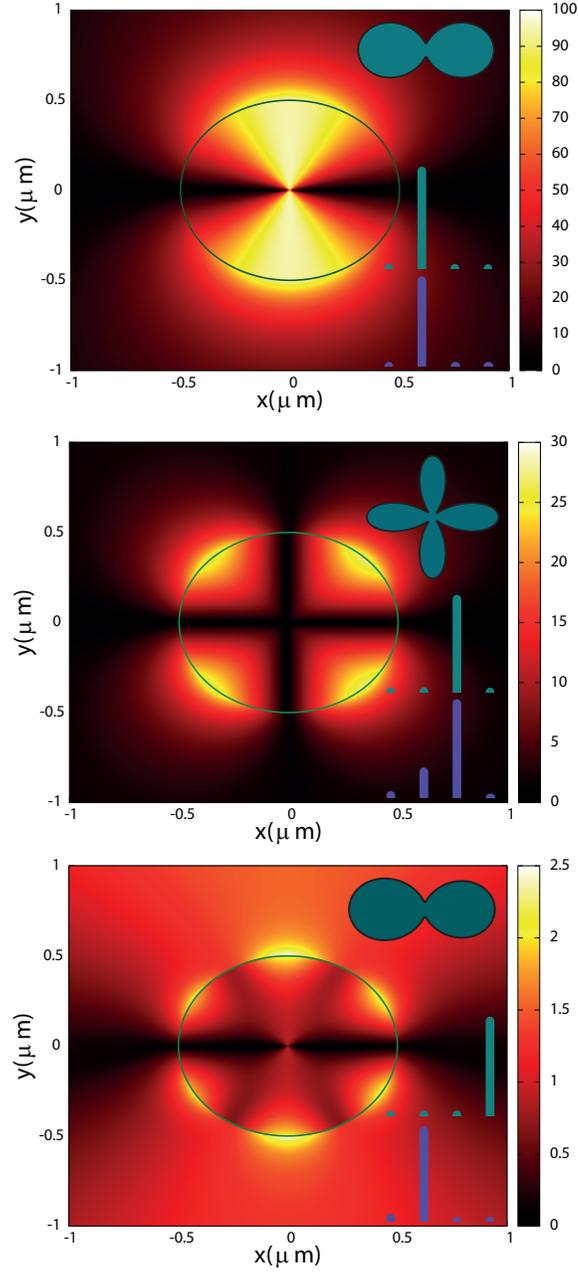}}
\caption{Map of $|\vec E (\rho, \varphi)|$ for a graphene-coated cylinder illuminated by a p-polarized plane wave.  
The incident wavelength is $\lambda=28.809 \;\mu$m (top), $\lambda=20.295 \;\mu$m (center) and $\lambda=16.553 \;\mu$m (bottom). 
The chemical potential is $\mu_c=0.8$ eV and all other parameters are the same as in Figure~\ref{scat-absor-diel}. 
The circle (continuous line) indicates the graphene layer. The top inset illustrates schematically the angular distribution of scattered power, the  medium inset illustrates the values of $|c_n|$, $n=0,\, 1,\, 2,\, 3$ (the multipole coefficients inside the cilynder) and the bottom inset illustrates the values of $|a_n|$), $n=0,\, 1,\, 2,\, 3$ (the multipole coefficients outside the cilynder). }
\label{multipolos1}
\end{figure}

In Figure~\ref{multipolos1}a we plot the spatial distribution of the electric field normalized to the incident amplitude  for the wire considered in Figure~\ref{scat-absor-diel} and $\mu_c=0.8$ eV. The incident wavelength is 
$\lambda=28.809 \;\mu$m, a value equal to the real part of the complex pole $\Lambda_1$ (see Table 1) of the multipole coefficients $a_1$ and $c_1$ and for which the strongest maxima in the scattering and absorption cross-sections for 
$\mu_c=0.8$ eV occur.
We observe that the picture is very similar to that obtained for subwavelength metallic wires:  the near fields are clearly of an electric dipole nature, with the dipolar moment oriented in the direction of the incident electric field ($\hat y$).  
However, contrary to the case of subwavelength metallic (impenetrable) wires, where the field cannot penetrate into the interior regions and is limited to a surface layer of approximately one skin depth thick, in the graphene-coated dielectric wire there are regions where the interior field can be greatly enhanced. 
Taking into account that 
the electric field induced in the scatterer is much stronger than that of incident radiation (see scale in Figure~\ref{multipolos1}a), the far-field intensities are dominated by an electric dipole pattern, as indicated in the inset in Figure~\ref{multipolos1}a, 
where we have sketched the angular distribution of scattered power for this excitation wavelength. 
When the incident wavelength is $\lambda=20.295 \;\mu$m, the scattering cross-section for $\mu_c=0.8$ eV (see Figure~\ref{scat-absor-diel}) exhibits a very weak peak while the absorption cross-section for the same value of $\mu_c$ exhibits a strong enhancement (an enhancement factor near three orders of magnitude greater than the non-resonant case). 
At this wavelength, the electric field induced in the scatterer, although not as strong, is again stronger than that of incident radiation and of a quadrupolar, not dipolar, nature. This is shown in Figure~\ref{multipolos1}b, where we plot the map of  $|\vec E (\rho, \varphi)|$  normalized to the incident amplitude  and calculated for the same parameters as in Figure~\ref{multipolos1}a, except the incident wavelength 
$\lambda=20.295 \;\mu$m. Since the electric field induced in the scatterer is much stronger than that of incident radiation, the far-field intensities are dominated by an electric quadrupole pattern, as indicated in the inset in Figure~\ref{multipolos1}a. 
In Figure~\ref{multipolos1}c we show the map of  $|\vec E (\rho, \varphi)|$, normalized to the incident amplitude, for the same parameters as in Figure~\ref{multipolos1}a and \ref{multipolos1}b, except that now the incident wavelength is changed to 
$\lambda=16.553 \;\mu$m. For this value of $\lambda$ and for $\mu_c=0.8$ eV a weak peak can be observed in the absorption cross-section, although no peak is observed in the scattering cross-section. 
We observe that the spatial distribution of the electric field induced in the scatterer at this excitation wavelength is hexapolar-shaped. 
However, the far-field intensities are dominated by an electric dipole pattern, as indicated in the bottom schematic diagram 
in Figure~\ref{multipolos1}c) which illustrates the values of $|a_n|$ and $|c_n|$, $n=0,\, 1,\, 2,\, 3$. 
Similar results have been obtained for the other  values of chemical potential considered in Figure~\ref{scat-absor-diel}. 
\begin{figure}[htbp!]
\centering
\resizebox{0.85\textwidth}{!}
{\includegraphics{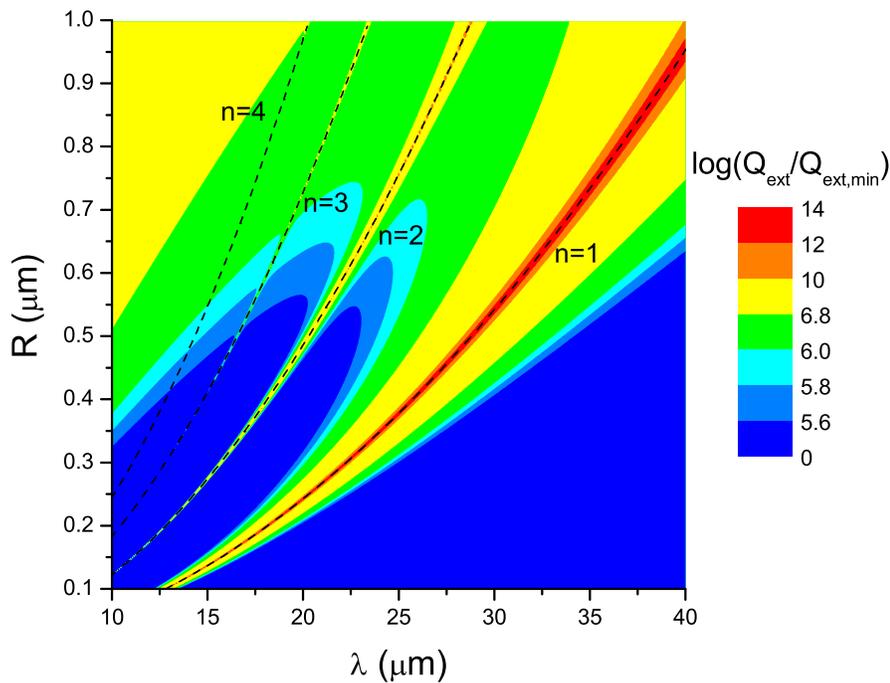}}
\caption{Extinction cross-section in a graphene-coated dielectric wire in vacuum ($\mu_{2}=\varepsilon_{2}=1$), explored as a function of the wire radius $R$ and the excitation wavelength $\lambda$. 
The dashed lines indicate the resonance position, numerically obtained as the poles of the coefficients $a_n$. 
The incident wave is $p$-polarized, $\mu_c=0.8$ eV, $\varepsilon_{1}=3.9$ and $\mu_{1}=1$. } 
\label{dispersionRadio}
\end{figure}

To investigate the size dispersion of the localized plasmonic modes of graphene-coated wires, we show in Figure~\ref{dispersionRadio} a color map of the extinction cross-section in the $R-\lambda$ plane. 
The incident wave is $p$-polarized, $\mu_c=0.8$ eV, the ambient medium is vacuum, $\varepsilon_{1}=3.9$ and $\mu_{1}=1$.  
We observe that the enhancements in the extinction cross-section follow the dashed lines labeled with $n$. 
These lines indicate the resonance position, numerically obtained as the real part of the pole of the coefficient $a_n$ given by equation (\ref{an}). 

\begin{figure}[htbp!]
\vspace{1cm}
\centering
\resizebox{0.80\textwidth}{!}
{\includegraphics{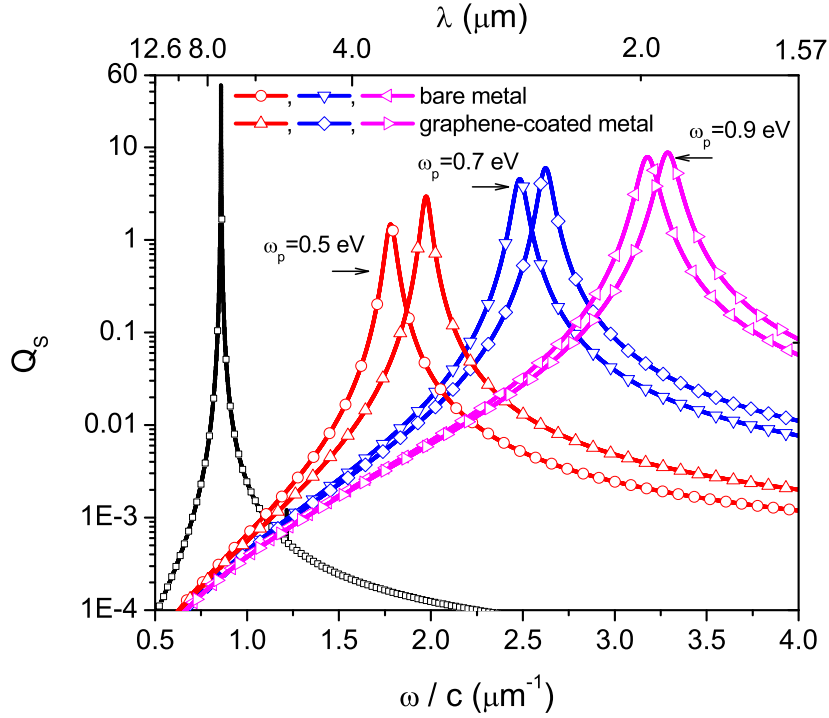}}
\caption{Scattering cross-section spectrum in vacuum for a graphene-coated metallic wire with a radius $R=50 \;$nm illuminated under p polarization. $\sigma(\omega)$ is given by the Kubo expression with $T=300^\circ$ K, $\gamma_c=0.1$ meV and $\mu_c=0.5$eV. $\varepsilon_1(\omega)$ is given by equation (\ref{drude1}), with $\varepsilon_{\infty}=1$, $\gamma_{m}=0.01$ eV and 
$\hbar \omega_p=$ 0.5eV, 0.7 and 0.9eV. The scattering cross-sections for the bare metallic cylinders 
and for an empty graphene cylinder in vacuum (black curve, $\varepsilon_{1}=\varepsilon_{2}=1$ and $\mu_{1}=\mu_{2}=1$) 
are also given as a reference.
}
\label{metal1}
\end{figure}

The examples so far highlight the attractive plasmonic features that a graphene coating can produce in intrinsically  nonplasmonic wires. Next, we consider graphene-coated wires made of metallic or metallic-like substrates, that is, 
intrinsically plasmonic wires, where the localized surface plasmons of the graphene coating are expected to modify the localized plasmons already existing in the bare particle. 
We assume that the dispersive behavior of the interior electric permittivity is described by the Drude model 
\begin{equation}
\varepsilon_1(\omega)=\varepsilon_{\infty}-\frac{\omega_p^2}{\omega^2+i\gamma_{m}\omega}\,,
\label{drude1}
\end{equation}
with $\varepsilon_{\infty}$ the residual high-frequency response of the material, 
$\omega_p$ the metallic plasma frequency and $\gamma_{m}$ the optical loss-rate of the Drude material. 
This model is applied to describe strongly doped semiconductors \cite{drudematerial1} which 
allow dynamic manipulation of carrier densities and have plasma frequencies in the same range 
that the realizable Fermi energies of graphene. 
Here the situation is similar to that reported for graphene-coated Drude spheres \cite{scatsphere} in the sense that the region of induced charges in the graphene coating is not separated from the region of induced charges in the metallic substrate. For the case considered in this paper, both regions coincide with the cylindrical surface of radius $R$ and therefore 
a single hybridized plasmon is expected, rather than two (bonding and antibonding) hybridized plasmons with different energies,  characteristic of systems where induced charges occur in two separated regions. 
In Figure~\ref{metal1} we plot the scattering cross-section spectrum in vacuum for a graphene-coated metallic wire with a radius $R=50 \;$nm illuminated under p polarization. The interior electric permittivity $\varepsilon_1(\omega)$ is given by equation (\ref{drude1}) with $\varepsilon_{\infty}=1$, $\gamma_{m}=0.01$ eV and different values of $\hbar \omega_p$ (0.5 eV, 0.7 eV and 0.9 eV), while the values of the complex surface conductivity of graphene $\sigma(\omega)$ have been obtained from the Kubo expression with $T=300^\circ$ K, $\gamma_c=0.1$ meV and $\mu_c=0.5$eV. The scattering cross-sections for the bare metallic cylinders and for 
an empty graphene cylinder in vacuum (black curve, $\varepsilon_{1}=\varepsilon_{2}=1$ and $\mu_{1}=\mu_{2}=1$) 
are also given as a reference. 
The excitation frequencies are in the range between $23.87$ THz (incident wavelength $12.57 \;\mu m$) and $191$ THz (incident wavelength $1.57 \;\mu$m). 

\begin{figure}[htbp!]
\vspace{1cm}
\centering
\resizebox{0.80\textwidth}{!}
{\includegraphics{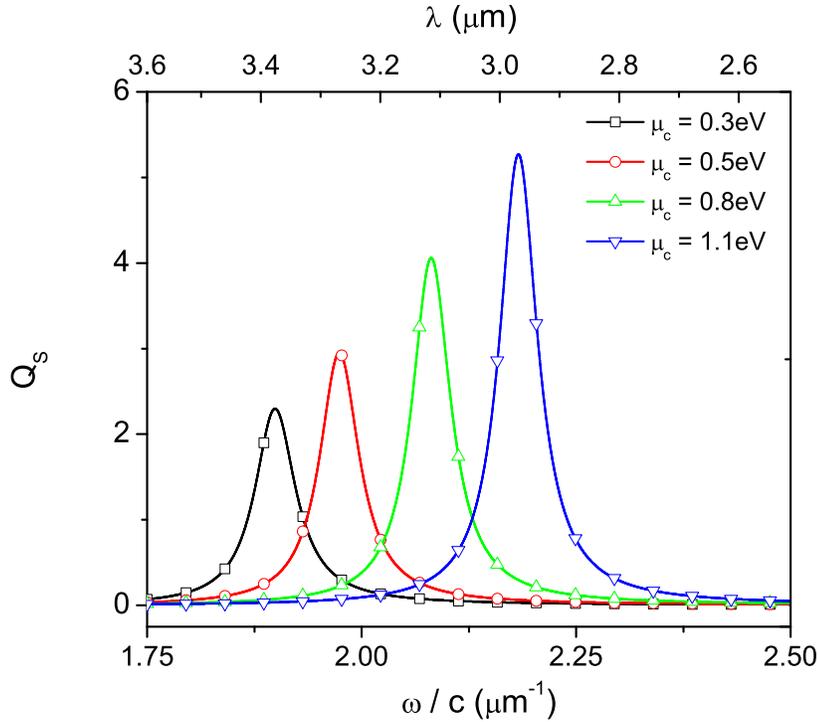}}
\caption{Scattering cross-section spectrum in vacuum for a graphene-coated metallic wire ($\hbar \omega_p=$ 0.7eV) 
illuminated under p polarization for different values of the chemical potential $\mu_c$ (0.3eV, 0.5, 0.8 and 1.1 eV). }
\label{metal2}
\end{figure}

The curves in Figure~\ref{metal1} show that the hybridized resonances are always blueshifted compared to the resonances of the bare wire. This is consistent with the fact that the net effect of the graphene coating is to increase the induced charge  density on the cylindrical surface of the metallic wire. The hybridized resonances can be controlled with the help of a gate voltage, which ultimately determines the value of the chemical potential $\mu_c$. To explore the tunability of the hybridized resonances, in Figure~\ref{metal2} we give scattering cross-section spectra for a graphene-coated metallic wire ($\hbar \omega_p=$ 0.5eV) illuminated under p polarization for different values of the chemical potential $\mu_c$ (0.3eV, 0.5, 0.8 and 1.1 eV). 

\section{Summary} \label{conclusiones}

Fueled by recent experiments \cite{exp1,exp2,exp3,exp4} showing that, thanks to the van der Waals force, a graphene sheet can be tightly coated on a fiber surface, in this paper we have investigated the electromagnetic response of circular cross--section wires coated with a graphene monolayer. The emphasis has been centered around the plasmonic properties of the coated wire, a system that not only is interesting for several applications, but provides a simple canonical model for the plasmonic characteristics of other graphene--coated 2D particles. 
We developed an analytical method of scattering based on the separation of variables approach and obtained 
a rigorous solution in the form of an infinite series of cylindrical multipole partial waves. 
We presented examples for both dielectric (nonplasmonic) and metallic (plasmonic) wire substrates. In the first case the graphene coating introduces surface plasmons mechanisms which were absent in the bare particle. In the second case, there is an interaction between the plasmons supported independently by the bare metallic wire and those suported by the graphene coating. In both cases, the examples show that the good tunability of the graphene coating leads to unprecedented control over the location and magnitude of the particle plasmonic resonances.

\section*{Acknowledgments}
The authors acknowledge the financial support of Consejo Nacional de Investigaciones Cient\'{\i}ficas y T\'ecnicas, (CONICET, PIP 1800)
and Universidad de Buenos Aires (project UBA 20020100100327).


\end{document}